\documentclass[conference]{IEEEtran}
\IEEEoverridecommandlockouts

\usepackage{cite}
\usepackage{amsmath,amssymb,amsfonts}
\usepackage{algorithmic}
\usepackage{graphicx}
\usepackage{textcomp}
\usepackage{xcolor}
\usepackage{booktabs}
\usepackage{url}
\usepackage{balance}
\def\BibTeX{{\rm B\kern-.05em{\sc i\kern-.025em b}\kern-.08em
    T\kern-.1667em\lower.7ex\hbox{E}\kern-.125emX}}

\newcommand{\coauth}{\texttt{Co-Authored-By}}
\newcommand{\pp}{\,pp}
\newcommand{\auc}{AUC}

\begin{document}

\title{Reliable Mining of Heterogeneous\\ AI Coding-Agent Data: Context-Aware Prediction\\ and Resource-Aware Ranking}

\author{\IEEEauthorblockN{Haoran Yu}
\IEEEauthorblockA{\textit{Independent Researcher}\\
United States\\
haoranyu889@gmail.com}
\and
\IEEEauthorblockN{Lifei Liu}
\IEEEauthorblockA{\textit{Independent Researcher}\\
United States\\
lliu.lifei@gmail.com}
\and
\IEEEauthorblockN{Danping Zhang}
\IEEEauthorblockA{\textit{Nanchang Hangkong University}\\
China\\
zhangdanping@nchu.edu.cn}
}

\maketitle

\begin{abstract}
Operational records from AI coding agents differ across agents, repositories, tasks, and workflows. We study 71{,}677 pull requests (PRs) from six agents and show that pooling these records can reverse descriptive and predictive conclusions. PRs carrying a co-authorship trailer have a 55.6\% merge rate, compared with 77.5\% without the trailer, but removing the dominant agent flips this $-21.8$ percentage-point gap to $+23.7$ points. We treat this Simpson-type reversal as a data-veracity problem and represent source, environment, and workflow context in a feature cascade. Gradient-boosting merge-prediction \auc{} increases from 0.774 to 0.832; the advantage persists under temporal evaluation (0.801) and unseen repositories (0.807). The co-authorship coefficient changes from $-1.09$ to $+1.44$ after conditioning, whereas repository-clustered AIPW intervals show no stable adjusted trailer association. A strict PR-creation-time model reaches 0.741 \auc{} in standard evaluation and 0.710 on unseen repositories, although its temporal \auc{} falls to 0.641. Finally, a separately cross-fitted analysis of an observed-mode functional estimates a $+12.3$-point doubly robust gain at a 30\% allocation budget on the non-Codex population, only $0.5$ points above risk ranking. This last result is retrospective decision support over metadata states, not a causal effect of assigning human review.
\end{abstract}


\section{Introduction}

AI coding agents (GitHub Copilot, OpenAI Codex, Devin, Cursor, Claude Code, and Google Jules) now draft, commit, and open pull requests (PRs) in real repositories~\cite{jimenez2024,hou2024}. Their operation leaves structured records of commits, PR metadata, repository interactions, and author/committer attribution. These records support mining questions about which PRs succeed and which contexts matter. They are also heterogeneous: different agents operate in different repositories, on different tasks, and under different development workflows.

Classical pooled analysis can obscure differences in the processes that generated the observations. Using the \coauth{} trailer as a workflow marker and pooling all 71{,}677 PRs from six agents, trailer-bearing PRs merge at 55.6\%, compared with 77.5\% for PRs without the trailer, a $-21.8$ percentage-point (\pp{}) gap. A model trained on the pooled data learns the same negative co-authorship coefficient.

\begin{figure}[t]
\centering
\includegraphics[width=\columnwidth]{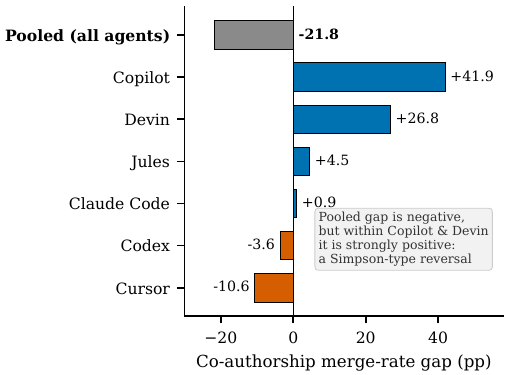}
\caption{Pooled and within-agent co-authorship gaps differ. Across all agents, trailer-bearing PRs merge $21.8\pp$ less often than PRs without the trailer, while Copilot and Devin show positive within-agent gaps. The reversal arises from agent composition.}
\label{fig:paradox}
\end{figure}

This is a Simpson-type compositional reversal~\cite{simpson1951,pearl2014} (Fig.~\ref{fig:paradox}). Codex contributes 49.6\% of the data, merges at a high rate (82.9\%), and uses the trailer in only 1.2\% of its PRs; removing Codex flips the pooled gap to $+23.7\pp{}$. We treat the reversal as a \emph{data-veracity} problem because the pooled statistic does not describe the within-agent pattern and its sign is inherited by a predictor trained without source context.

Our response is to make data-generating context explicit. We define a \emph{context cascade} that adds source context (agent), environment context (historical repository activity), and workflow context (task, PR structure, and observed author/committer attribution mode). We use the cascade to explain the reversal, improve retrospective merge prediction, and study resource-aware ranking. The attribution mode is derived from names in commit metadata and is not a validated measure of human effort. We therefore report the decision result as retrospective ranking over observed modes, not as the effect of assigning human review.

\noindent\textbf{Contributions.}
\begin{itemize}
\item \textbf{Heterogeneous-data reliability.} In 71{,}677 PRs from six agents, we show that agent composition reverses the aggregate co-authorship association, while omitted source, repository, and workflow context similarly reverses the coefficient learned by a pooled predictive model.
\item \textbf{Context-aware predictive mining.} We develop a source/environment/workflow cascade that raises gradient-boosting merge-prediction \auc{} from 0.774 to 0.832 after correcting repository activity to use only prior PRs. The advantage persists under temporal and unseen-repository evaluation, and the co-authorship coefficient changes from $-1.09$ to $+1.44$ after conditioning.
\item \textbf{Resource-aware decision support.} Using only PR-creation-time covariates for ranking, we evaluate a 30\% allocation over observed attribution modes on the non-Codex population. The cross-sectional doubly robust point estimate is $+12.3\pp{}$ over the bot-attributed baseline (repository-clustered 95\% CI $[+10.7,+14.1]$); temporal evaluation retains a $+10.3\pp{}$ point estimate but no resolved advantage over risk ranking. This is a statistical functional of observed metadata modes, not a human-review intervention.
\end{itemize}

The remainder of the paper is organized as follows. Section~\ref{sec:related} surveys related work; Section~\ref{sec:data} describes the data and problem; Section~\ref{sec:framework} defines the context-aware mining framework; Sections~\ref{sec:bias}--\ref{sec:routing} present bias discovery, predictive mining, and resource-aware routing; Section~\ref{sec:robustness} reports robustness to distribution change; Sections~\ref{sec:discussion}--\ref{sec:conclusion} discuss threats and conclude.

\section{Related Work}
\label{sec:related}

\paragraph{Heterogeneous and context-aware data mining}
Simpson's paradox~\cite{simpson1951}, the reversal of an association after conditioning on a grouping variable, is the classic symptom of mining a heterogeneous mixture as if it were homogeneous, with the Berkeley admissions case~\cite{bickel1975} the canonical example. Pearl~\cite{pearl2014,pearl2009} and Greenland et al.~\cite{greenland1999} formalize when the within-stratum association is the appropriate lens. Our setting differs from these textbook illustrations in two ways that matter for data mining: the mixture components (agents, repositories, workflows) are numerous and imbalanced rather than a single binary confounder, and the reversal is not merely descriptive because predictive models trained on pooled data inherit its direction. We therefore treat compositional heterogeneity as a reliability property of the mining pipeline.

\paragraph{Mining AI-agent operational data}
LLM-based coding agents have been evaluated on code-generation benchmarks~\cite{jimenez2024}, code quality~\cite{dakhel2023,nguyen2022}, and developer productivity~\cite{ziegler2024,peng2023,mozannar2024}, and surveyed broadly~\cite{hou2024}. The AIDev dataset~\cite{aidev} exposes PR-level activity across agents. Recent empirical studies examine agentic-PR usage~\cite{watanabe2025agentic,mazloomzadeh2026contribute}, failed PRs~\cite{ehsani2026fail}, and unmerged fix-related PRs~\cite{alam2026unmerged}. These studies establish that agentic PRs vary in outcome and workflow. Our analysis asks a different reliability question: how cross-agent composition changes pooled conclusions and learned models, and whether explicit context improves prediction under repository, time, and agent shift.

\paragraph{Pull-request and repository analytics}
Merge outcomes depend on technical factors (PR size and complexity) and social factors (contributor reputation and reviewer workload)~\cite{gousios2015,yu2016,tsay2014}; observational PR data do not by themselves identify causal effects~\cite{vasilescu2015}. A parallel line studies how automation bots reshape PR workflows~\cite{wessel2018,wessel2022,soares2022}. Closest to our predictive analysis, Pansuriya et al.~\cite{pansuriya2026} use AIDev to predict PR acceptance and review effort from submission-time features. Peralta et al.~\cite{peralta2026} study why agentic PRs are merged or rejected, and Xu et al.~\cite{xu2026agenticpr} quantify their frequency, structure, and merge-conflict rates. Our focus is complementary: we analyze composition-induced descriptive and coefficient reversals, then evaluate context representations under repository, temporal, agent, and mixture shifts; strict creation-time prediction is retained as a boundary audit rather than treated as evidence of deployment readiness.

\paragraph{Resource-aware and uplift-based decisions}
Turning predictions into decisions under a budget connects to uplift modeling and heterogeneous-treatment-effect estimation: metalearners~\cite{kunzel2019}, causal forests~\cite{athey2016,wager2018}, and uplift surveys~\cite{gutierrez2017} rank units by an estimated outcome contrast rather than by outcome level. Doubly robust adjustment~\cite{hernan2024} combines propensity and outcome models. We adapt this machinery to estimate a statistical functional over observed metadata modes. Because attribution mode is observational and is not an assigned review action, we do not interpret the resulting value as a causal human-review effect.

\section{Data and Problem Formulation}
\label{sec:data}

\subsection{AIDev Repository Data}
We use an expanded AIDev snapshot~\cite{aidev}, downloaded from the public dataset repository on August 30, 2026.\footnote{\url{https://huggingface.co/datasets/hao-li/AIDev}} We work from its curated per-PR record table rather than the raw agent-PR firehose. This release yields $N=71{,}677$ PRs from six agents across 6{,}673 distinct repositories, spanning 2024-12 to 2025-10; it is larger than the five-agent snapshot described in the original dataset paper. Each PR is linked to its hosting repository and final state. Merge status is a non-null merge timestamp; open PRs are treated conservatively as ``not merged,'' and Section~\ref{sec:robustness} reports a censoring check. Repositories are distinct \texttt{repo\_id} values. Because a repository can host several agents, the per-agent counts in Table~\ref{tab:composition} sum to more than 6{,}673.

Table~\ref{tab:composition} summarizes the composition. Codex contributes 49.6\% of all PRs, making its workflow profile a major determinant of pooled statistics (Section~\ref{sec:bias}).

\begin{table}[t]
\caption{Dataset composition by AI coding agent. Codex contributes 49.6\% of PRs. Repository counts do not sum to 6{,}673 because repositories can host multiple agents.}
\label{tab:composition}
\centering
\begin{tabular}{lrrrr}
\toprule
\textbf{Agent} & \textbf{$n$ PRs} & \textbf{\%} & \textbf{Repos} & \textbf{MR} \\
\midrule
Codex        & 35{,}521 & 49.6 & 1{,}918 & 82.9\% \\
Copilot      & 23{,}496 & 32.8 & 3{,}791 & 50.8\% \\
Devin        & 6{,}172  & 8.6  & 336     & 55.1\% \\
Cursor       & 4{,}059  & 5.7  & 689     & 52.4\% \\
Claude Code  & 1{,}942  & 2.7  & 680     & 65.4\% \\
Google Jules & 487      & 0.7  & 132     & 58.1\% \\
\midrule
Total        & 71{,}677 & 100  & 6{,}673 & 67.6\% \\
\bottomrule
\end{tabular}
\end{table}

\subsection{Variety, Veracity, and Value}
This work focuses on the variety, veracity, and value dimensions of Big Data.
\emph{Variety} enters through six agents, thousands of repositories, several task types, and different trailer and attribution patterns. Agent observation windows also differ (Devin appears earliest, Codex latest), so pooled counts mix non-aligned time ranges.
\emph{Veracity} is the resulting measurement problem: merge rates and trailer frequencies differ systematically by agent, so a pooled association can disagree with within-agent patterns (Section~\ref{sec:bias}). \emph{Value} is addressed in Section~\ref{sec:routing}, where mined context supports budget-constrained retrospective ranking.

\subsection{Outcomes and Context Variables}
\paragraph{Outcome} The binary target is \emph{merge}: whether a PR reached the merged state.

\paragraph{Co-authorship trailer ($C$)} A PR has $C=1$ if any commit contains a \coauth{} trailer and $C=0$ otherwise. In total, 32{,}418 PRs (45.2\%) carry the trailer. Because agents may insert it themselves and its use is not standardized, it is a noisy workflow marker rather than ground truth for human effort.

\paragraph{Observed author/committer attribution mode} We aggregate identities over all commits in a PR and derive four name-based metadata states. At least one unmatched author and committer gives \texttt{human\_both} ($n=47{,}586$, MR 80.6\%); bot-only authors with at least one unmatched committer give \texttt{agent\_draft} ($n=851$, MR 58.0\%); and nonempty bot-only author and committer sets give \texttt{fully\_autonomous} ($n=23{,}046$, MR 41.3\%). The remaining 194 PRs have missing or asymmetric identities and are marked \texttt{unresolved}, rather than assigned to a bot-attributed state. Bot matching uses \texttt{[bot]}, \texttt{github-actions}, \texttt{dependabot}, \texttt{copilot}, \texttt{devin-ai}, \texttt{codex}, \texttt{cursor}, \texttt{claude}, or \texttt{jules} (case-insensitive); other names are called human-attributed. This heuristic has not been manually validated, some agents use human-looking identities, and the data contain no record of actual human effort. We therefore interpret attribution mode only as a full-lifecycle metadata state.

\paragraph{PR structural features ($S$)} We use commit count, distinct author and committer counts, and log change size. Change size sums commit-level additions and deletions after deduplicating the file-level details table by commit SHA.

\paragraph{Task type ($T$)} The current AIDev release no longer ships a task-type table, so we re-derive task type from the PR title using conventional-commit prefixes (\texttt{feat}, \texttt{fix}, \texttt{docs}, \texttt{test}, \texttt{refactor}, \dots) with a keyword fallback: feature ($n=24{,}850$), bug fix ($n=14{,}635$), documentation ($n=6{,}110$), refactoring ($n=3{,}833$), test ($n=1{,}062$), and other ($n=21{,}187$); the six buckets sum to 71{,}677. This is a transparent heuristic re-derivation rather than curated labels, and task type is a minor feature in what follows (Section~\ref{sec:prediction}).

\paragraph{Repository context ($R$)} Historical repository activity is the log-transformed number of PRs created in the repository before the focal PR. PRs with the same creation timestamp receive the same prior count. This avoids using future PRs while providing a limited proxy for repository maturity and engagement.

\subsection{Research Questions}
\begin{itemize}
\item \textbf{RQ1 (compositional bias).} How does heterogeneous agent composition distort aggregate patterns in AI-agent repository data?
\item \textbf{RQ2 (context-aware knowledge discovery).} Which agent, repository, task, and PR-structure contexts explain the misleading aggregate association?
\item \textbf{RQ3 (predictive mining).} Does context-aware feature modeling improve merge-outcome prediction across heterogeneous contexts and under distribution shift?
\item \textbf{RQ4 (resource-aware decisions).} Can context-aware estimates rank PRs under a fixed observed-attribution-mode allocation budget?
\end{itemize}

\section{Context-Aware Mining Framework}
\label{sec:framework}

\begin{figure*}[t]
\centering
\includegraphics[width=0.98\textwidth]{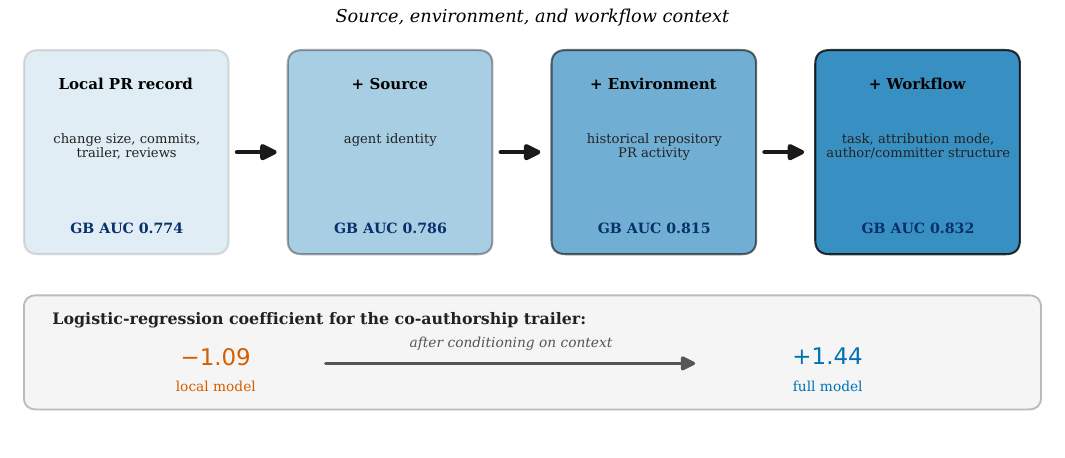}
\caption{The context cascade adds source, environment, and workflow variables to a local PR representation. With historical repository activity, gradient-boosting \auc{} rises from 0.774 to 0.832. The logistic-regression co-authorship coefficient changes from $-1.09$ in the local model to $+1.44$ after conditioning; this is a conditional predictive coefficient, not a causal effect.}
\label{fig:cascade}
\end{figure*}

We frame reliable mining of heterogeneous agent data as making data-generating context explicit in the feature representation. Rather than adding features ad hoc, we organize them into a \emph{context cascade} with a fixed semantic hierarchy, so that each level answers a specific source of heterogeneity (Fig.~\ref{fig:cascade}).

\subsection{Data Representation and the Context Cascade}
Let a PR be described by local features plus three context blocks. We construct four nested feature sets $X_0 \subset X_1 \subset X_2 \subset X_3$:
\begin{itemize}
\item \textbf{$X_0$, context-free (local).} PR-local signals: log change size, log commit count, the co-authorship indicator, and review count (a review-derived signal, dropped in the review-signal-free variant of Section~\ref{sec:prediction}). This is the naive mining baseline.
\item \textbf{$X_1$, \emph{source} context.} Adds agent identity, addressing source-mixture heterogeneity (which autonomous system produced the PR).
\item \textbf{$X_2$, \emph{environment} context.} Adds historical repository activity, computed from PRs created before the focal PR.
\item \textbf{$X_3$, \emph{workflow} context.} Adds observed attribution mode, task type, a changes-requested flag, and the numbers of distinct authors and committers. The changes-requested flag and review count are excluded from the review-signal-free variant.
\end{itemize}
The cascade is thus a structured heterogeneous-data representation, not merely a larger feature list: each level is a proper superset of the previous one and targets a named heterogeneity axis.

\subsection{Predictive Models}
We evaluate the cascade across logistic regression (LR), gradient boosting (GB), XGBoost~\cite{chen2016xgboost}, and LightGBM~\cite{ke2017lightgbm}. Nominal variables (agent, task type, and attribution mode) are one-hot encoded within each training fold. All learners use fixed, untuned configurations so that the comparison isolates feature representation rather than per-model tuning. The full workflow level contains 11 raw variable groups; one-hot encoding expands the fitted design matrix beyond 11 columns.

\subsection{Evaluation Protocols}
To separate genuine context signal from memorization, we evaluate under three protocols:
\begin{enumerate}
\item \textbf{Standard cross-validation} (5-fold stratified) for in-distribution performance.
\item \textbf{Temporal split}: train on PRs before 15 August 2025, test on PRs after, measuring generalization to future data.
\item \textbf{Unseen-repository split}: grouped $k$-fold with the repository as the grouping unit, so no repository appears in both train and test, measuring transfer to new repositories rather than recall of repository identity.
\end{enumerate}
We report ROC-\auc{} as the primary metric and Brier score as a calibration-sensitive secondary metric. Because repository context is one of the strongest predictors, the contrast between standard and unseen-repository CV tests whether the model transfers beyond repositories seen during training.

\subsection{Feature Availability}
The full cascade is a retrospective representation: review count, changes requested, commit structure, change size, and observed attribution mode can reflect events after a PR is opened. We report a \emph{review-signal-free retrospective} variant that removes the two explicit review variables, but do not label it point-in-time. Separately, a strict \emph{creation-time} variant uses only agent identity, title-derived task type, and historical repository activity. The decision analysis in Section~\ref{sec:routing} uses these creation-time covariates for ranking, while treating full-lifecycle attribution mode only as the observed metadata state being contrasted.

\section{Discovering Composition-Driven Bias}
\label{sec:bias}

This section answers RQ1 and RQ2: how agent composition distorts aggregate patterns, and which contexts explain the distortion.

\subsection{The Pooled Reversal (RQ1)}
Table~\ref{tab:pooled} shows the aggregate result. Trailer-bearing PRs merge at 55.6\% and PRs without the trailer at 77.5\%, a $-21.8\pp{}$ gap. Table~\ref{tab:peragent} stratifies by agent and changes the pattern: Copilot ($+41.9\pp$) and Devin ($+26.8\pp$) show positive within-agent gaps, while Codex and Jules, which contribute heavily to the pooled no-trailer rate, show smaller gaps ($-3.6$ and $+4.5\pp$).

\begin{table}[t]
\caption{Pooled merge rates by co-authorship status. The $-21.8\pp{}$ gap is confounded by agent composition.}
\label{tab:pooled}
\centering
\begin{tabular}{lrr}
\toprule
\textbf{\coauth{}} & \textbf{$n$ PRs} & \textbf{Merge rate} \\
\midrule
Present & 32{,}418 & 55.6\% \\
Absent  & 39{,}259 & 77.5\% \\
\midrule
Delta & \multicolumn{2}{r}{$-21.8\pp{}$} \\
\bottomrule
\end{tabular}
\end{table}

\begin{table}[t]
\caption{Per-agent merge rates by trailer status. Codex and Jules use the trailer in $\approx$1--3\% of PRs and contribute heavily to the pooled no-trailer rate.}
\label{tab:peragent}
\centering
\begin{tabular}{lrrrr}
\toprule
\textbf{Agent} & \textbf{\% trailer} & \textbf{MR(present)} & \textbf{MR(absent)} & \textbf{$\Delta$} \\
\midrule
Codex        & 1.2\%  & 79.3\% & 82.9\% & $-3.6\pp$ \\
Copilot      & 89.8\% & 55.1\% & 13.2\% & $+41.9\pp$ \\
Devin        & 95.7\% & 56.2\% & 29.4\% & $+26.8\pp$ \\
Cursor       & 84.6\% & 50.8\% & 61.4\% & $-10.6\pp$ \\
Claude Code  & 78.4\% & 65.6\% & 64.8\% & $+0.9\pp$ \\
Google Jules & 3.3\%  & 62.5\% & 58.0\% & $+4.5\pp$ \\
\bottomrule
\end{tabular}
\end{table}

\subsection{Agent Composition as the Mechanism}
Codex contributes 49.6\% of all PRs, has an 82.9\% merge rate among no-trailer PRs, and uses the trailer in only 1.2\% of submissions. This combination raises the pooled no-trailer rate; Jules has a smaller contribution in the same direction. Copilot and Devin use the trailer frequently and have lower overall merge rates. Removing Codex flips the pooled gap from $-21.8\pp{}$ to $+23.7\pp{}$. The reversal therefore depends strongly on agent mixture and does not identify an effect of co-authorship.

\subsection{Progressive Context Conditioning (RQ2)}
Conditioning further attenuates the large within-agent gaps. A repository$\times$agent fixed-effects linear model on the 1{,}142 repo--agent pairs (29{,}603 PRs) containing both PR types yields an overall within-pair association of $+20.9\pp{}$ ($p<0.001$). The estimate is concentrated in Copilot ($+35.4\pp{}$ within repositories), whose no-trailer PRs are mostly single-commit drafts. Adding commit count reduces the Copilot estimate to $+24.1\pp{}$; restricting to multi-commit PRs reduces it to $-1.1\pp{}$ ($p=0.78$). Devin's cross-sectional $+26.8\pp{}$ gap similarly falls to $+2.8\pp{}$ ($p=0.53$) within repositories. Source, environment, and PR structure therefore account for much of the pooled and within-agent variation.

\subsection{Association Robustness under Covariate Adjustment}
\label{sec:robust_assoc}
We apply a cross-fitted augmented inverse-probability-weighting (AIPW) estimator~\cite{hernan2024}, adjusting for one-hot agent and task indicators, log change size, log commit count, and historical repository activity, with the \coauth{} trailer as the observational treatment. Co-authored and no-trailer PRs separate sharply by agent, and trimming to $[0.01,0.99]$ retains only 36{,}599 PRs. Table~\ref{tab:aipw} reports repository-clustered bootstrap intervals. The point estimate changes sign across trim levels and every interval spans zero. We therefore use AIPW only as an association-robustness check; it does not identify a causal trailer effect.

\begin{table}[t]
\caption{Cross-fitted AIPW association under propensity trimming, with repository-clustered bootstrap 95\% intervals.}
\label{tab:aipw}
\centering
\begin{tabular}{lrr}
\toprule
\textbf{Trim bounds} & \textbf{$n$ retained} & \textbf{Estimate (95\% CI)} \\
\midrule
$[0.01,0.99]$ & 36{,}599 & $-1.3\pp{}$ $[-5.6,+2.6]$ \\
$[0.05,0.95]$ & 18{,}893 & $+1.1\pp{}$ $[-1.9,+4.0]$ \\
$[0.10,0.90]$ & 11{,}678 & $+1.7\pp{}$ $[-0.9,+4.6]$ \\
\bottomrule
\end{tabular}
\end{table}

\section{Context-Aware Predictive Mining}
\label{sec:prediction}

We answer RQ3 by evaluating the cascade in retrospective, distribution-shift, review-signal-free, and strict creation-time settings.

\subsection{Cascade Prediction Results}
Table~\ref{tab:cascade} reports 5-fold stratified ROC-\auc{} with one-hot categorical encoding fit inside each training fold. All four learners improve monotonically as context is added. GB increases from 0.774 to 0.832, XGBoost~\cite{chen2016xgboost} from 0.772 to 0.834, LightGBM~\cite{ke2017lightgbm} from 0.771 to 0.834, and LR from 0.678 to 0.786. The full GB Brier score is 0.146. Environment context produces the largest GB increment, from 0.786 to 0.815.

\begin{table}[t]
\caption{Merge-prediction \auc{} by cascade level (5-fold CV; historical repository activity).}
\label{tab:cascade}
\centering
\begin{tabular}{lrrrr}
\toprule
\textbf{Feature set} & \textbf{LR} & \textbf{GB} & \textbf{XGB} & \textbf{LGBM} \\
\midrule
Local              & 0.678 & 0.774 & 0.772 & 0.771 \\
$+$ Source         & 0.736 & 0.786 & 0.785 & 0.784 \\
$+$ Environment    & 0.760 & 0.815 & 0.814 & 0.815 \\
$+$ Workflow       & 0.786 & 0.832 & 0.834 & 0.834 \\
\midrule
Total gain         & $+10.8\pp{}$ & $+5.8\pp{}$ & $+6.2\pp{}$ & $+6.4\pp{}$ \\
\bottomrule
\end{tabular}
\end{table}

\subsection{Coefficient Reversal and Feature Attribution}
In logistic regression, the co-authorship coefficient is $-1.09$ in the local model and $+1.44$ (odds ratio 4.22) in the full cascade, with agent, task, and attribution mode one-hot encoded. The sign change shows that omitted context can make a pooled predictive model learn the wrong direction of association; neither coefficient is interpreted causally.

We compute grouped permutation importance on held-out folds, avoiding integer-coded categorical variables and training-set impurity importance. Permuting observed attribution mode decreases \auc{} by 0.122 on average, followed by historical repository activity (0.071), review count (0.070), and change size (0.049). Agent identity decreases \auc{} by 0.008, task type by 0.005, and the trailer by 0.002 after conditioning. The attribution-mode importance is predictive, but its name-based construction prevents interpreting it as measured human effort.

\subsection{Temporal, Repository, and Feature-Availability Tests}
Table~\ref{tab:generalization} shows that the retrospective context advantage persists under temporal and unseen-repository evaluation. The temporal gain is $+2.6\pp{}$, while the unseen-repository gain is $+5.9\pp{}$. Removing review count and changes requested gives a \emph{review-signal-free retrospective} model, not a point-in-time model, because commit structure, change size, and attribution mode summarize the completed PR history.

\begin{table}[t]
\caption{GB generalization. RSF denotes the review-signal-free retrospective feature set.}
\label{tab:generalization}
\centering
\begin{tabular}{lrrr}
\toprule
\textbf{Protocol} & \textbf{Local} & \textbf{Full} & \textbf{$\Delta$} \\
\midrule
Standard 5-fold CV          & 0.774 & 0.832 & $+5.8\pp$ \\
Temporal                    & 0.775 & 0.801 & $+2.6\pp$ \\
Unseen repository           & 0.748 & 0.807 & $+5.9\pp$ \\
RSF, standard               & 0.730 & 0.802 & $+7.1\pp$ \\
RSF, unseen repository      & 0.704 & 0.776 & $+7.3\pp$ \\
\bottomrule
\end{tabular}
\end{table}

\subsection{Strict Creation-Time Audit}
For a deployment-oriented audit, we restrict the representation to fields available when the PR is opened: agent, title-derived task type, and the number of earlier PRs in the repository. This model reaches 0.741 \auc{} in standard CV, 0.710 on unseen repositories, and a Brier score of 0.185. Its temporal \auc{} is only 0.641. The context is therefore useful at creation time, but temporal deployment is substantially harder than the retrospective results suggest. We report this gap rather than treating removal of review signals as sufficient evidence of point-in-time validity.

\section{Resource-Aware Observed-Mode Ranking}
\label{sec:routing}

RQ4 asks whether context supports ranking under a fixed observed-attribution-mode allocation budget. This section does not model assignment of human review. It contrasts two full-lifecycle metadata states: bot-attributed author and committer (raw label \texttt{fully\_autonomous}) and human-attributed author and committer (\texttt{human\_both}). The 851 mixed-mode and 194 unresolved PRs are excluded.

\subsection{Budget-Constrained Formulation}
Let $A\in\{\text{bot-attributed},\text{human-attributed}\}$ denote the observed attribution mode, and define the conditional mean $m(\mathbf{x},a)=\mathbb{E}[M\mid\mathbf{X}=\mathbf{x},A=a]$. For a ranking rule $\pi$, we study the standardized observed-mode functional $V_{\mathrm{obs}}(\pi)=\mathbb{E}_{\mathbf{X}}[m(\mathbf{X},\pi(\mathbf{X}))]$. Under a budget $B$, the ranking problem is
\begin{equation}
\max_{\pi}\ V_{\mathrm{obs}}(\pi)
\quad\text{s.t.}\quad
P_{\mathbf{X}}\!\left(\pi(\mathbf{X})=\text{human-attributed}\right)\le B,
\label{eq:routing}
\end{equation}
where $\mathbf{X}$ contains only creation-time covariates: agent, title-derived task type, and historical repository activity. Equation~\eqref{eq:routing} is a statistical standardization over conditional associations in the observed data. It is not an interventional value because the data do not record assignment to these modes.

We fit an S-learner $\hat f(\mathbf{x},a)$ and rank PRs by
$u(\mathbf{x})=\hat f(\mathbf{x},\text{human-attributed})-\hat f(\mathbf{x},\text{bot-attributed})$.
The top-$B$ set is unchanged by calibration errors that preserve this ranking. This property does not provide robustness to arbitrary outcome-model misspecification.

\subsection{Doubly Robust Evaluation of the Observed-Mode Functional}
\label{sec:ope}
Codex has 35{,}498 human-attributed PRs and only 10 bot-attributed PRs after unresolved identities are removed, so full-population overlap fails. We therefore define the non-Codex population ($n=35{,}124$) as the primary target population. In this subset, raw estimated propensities remain within $[0.01,0.99]$; the context-aware importance-weight mass is 0.966 and effective sample size is 11{,}798.

Evaluation propensity and outcome models use 5-fold cross-fitting. Ranking scores are averaged over five additional cross-fits with different split seeds, and all learned rankings, including the risk baseline, are constructed from these folds rather than from the nuisance predictions used for evaluation. We report doubly robust (DR) estimates of $V_{\mathrm{obs}}$. The repository-clustered bootstrap interval is conditional on the fitted models.

At $B=0.30$, the cross-sectional context-aware ranking has an estimated DR gain of $+12.3\pp{}$ over always bot-attributed, with a 95\% interval of $[+10.7,+14.1]\pp{}$ (Table~\ref{tab:ope}). Its point estimate is $0.5\pp{}$ above risk ranking; this difference is not statistically resolved.

For temporal evaluation, we learn each ranking rule only from the 16{,}618 non-Codex PRs before August 15, 2025, then evaluate on the 18{,}506 later PRs. Evaluation nuisance functions are independently cross-fitted within the later sample, so a PR's outcome is not used to score itself. Raw evaluation propensities range from 0.141 to 0.969; for context-aware ranking, the clipped importance-weight mass is 1.015 and effective sample size is 5{,}992. Context-aware ranking yields $+10.3\pp{}$ ($[+8.8,+11.8]\pp{}$). Its point estimate is only $0.04\pp{}$ above risk ranking, with an unresolved interval of $[-0.9,+0.9]\pp{}$.

\begin{table}[t]
\caption{Non-Codex DR gain at a 30\% observed-mode allocation budget, relative to always bot-attributed.}
\label{tab:ope}
\centering
\begin{tabular}{lrr}
\toprule
\textbf{Ranking rule} & \textbf{Cross-sec.} & \textbf{Temporal} \\
\midrule
Random                         & $+9.6\pp$  & $+9.4\pp$ \\
Agent threshold                & $+10.5\pp$ & $+9.8\pp$ \\
Risk (lowest $\hat q_0$)       & $+11.8\pp$ & $+10.2\pp$ \\
Context-aware                  & $+12.3\pp$ & $+10.3\pp$ \\
\bottomrule
\end{tabular}
\end{table}

These estimates require adequate overlap and consistent nuisance estimation, and they standardize a name-derived, full-lifecycle metadata mode. Creation-time context supports retrospective ranking in both splits, but its advantage over risk ranking is small and unresolved. Neither result is evidence that assigning human review would produce the same gain.

\section{Robustness to Data-Distribution Change}
\label{sec:robustness}

Heterogeneous agent ecosystems are non-stationary: new agents enter (Google Jules appears only from mid-2025) and dominant agents' shares change. A mining pipeline for such data should degrade gracefully as the mixture shifts and as data scale changes. We test three axes.

\subsection{Agent-Mixture Shift}
The pooled co-authorship gap is a population property that changes from $-21.8\pp{}$ under the observed mixture to $-10.0\pp{}$ under equal agent shares and $+23.7\pp{}$ without Codex. Predictively, we fix one held-out test set ($n=21{,}504$, all agents) and vary only the training composition (Table~\ref{tab:mixture}). The full cascade changes from 0.832 under the observed training mixture to 0.823 under uniform agent sampling. Training without Codex reduces \auc{} on the common all-agent test set to 0.613. Reweighting represented agents has a modest effect, but transfer to an omitted dominant agent is poor.

\begin{table}[t]
\caption{Agent-mixture sensitivity, predictive. All rows are evaluated on one common held-out test set ($n=21{,}504$, all agents); only the training composition varies, so the \auc{}s are directly comparable.}
\label{tab:mixture}
\centering
\begin{tabular}{lr}
\toprule
\textbf{Training composition} & \textbf{\auc{} (common test)} \\
\midrule
Observed (as-is)      & 0.832 \\
Uniform (equal share) & 0.823 \\
Codex removed         & 0.613 \\
\bottomrule
\end{tabular}
\end{table}

\subsection{Unseen-Agent Generalization}
For each leave-one-agent-out (LOAO) run, we train on five agents and evaluate on the sixth; the unseen agent has no fitted one-hot column. The full cascade improves on the local model for five agents, with macro-average \auc{} increasing from 0.614 to 0.704 (Table~\ref{tab:loao}). Codex is the exception: its full-cascade \auc{} is 0.422 when trained only on the other agents. The result supports transfer for several agents but not for the dominant compositional outlier.

\begin{table}[t]
\caption{Leave-one-agent-out transfer: full cascade trained on five agents, tested on the held-out (unseen) agent. Transfer holds for five of six; Codex, the compositional outlier, is the exception.}
\label{tab:loao}
\centering
\begin{tabular}{lrr}
\toprule
\textbf{Held-out agent} & \textbf{Naive} & \textbf{Full} \\
\midrule
Google Jules & 0.447 & 0.801 \\
Copilot      & 0.725 & 0.798 \\
Devin        & 0.718 & 0.779 \\
Cursor       & 0.698 & 0.764 \\
Claude Code  & 0.596 & 0.660 \\
Codex        & 0.498 & 0.422 \\
\midrule
Macro-average & 0.614 & 0.704 \\
\bottomrule
\end{tabular}
\end{table}

\subsection{Data-Scale Sensitivity}
\begin{figure}[t]
\centering
\includegraphics[width=\columnwidth]{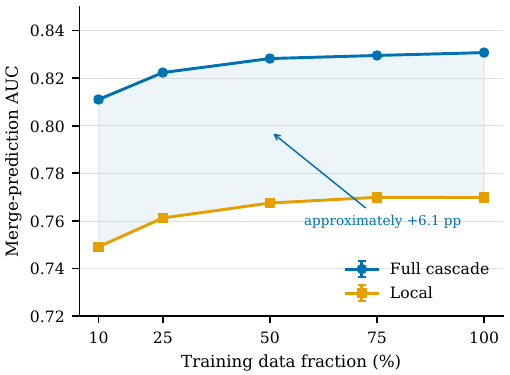}
\caption{Data-scale sensitivity on a fixed held-out set (mean $\pm$ standard deviation over three seeds). The context advantage remains approximately $6.1\pp{}$ across training fractions.}
\label{fig:scale}
\end{figure}

We train at increasing fractions of the data (10--100\%) over three seeds and evaluate on a fixed held-out set (Fig.~\ref{fig:scale}). The full-cascade advantage is stable at approximately $6.1\pp{}$: at 10\% of the training pool, \auc{} is 0.811 versus 0.749, and at 100\% it is 0.831 versus 0.770.

\subsection{Open-PR Censoring Sensitivity}
Treating open PRs (8.3\% of the data) as ``not merged'' can undercount eventual merges, especially for recent PRs. Restricting to resolved PRs preserves the pooled reversal ($-16.3\pp{}$ versus $-21.8\pp{}$ on the full data) and gives a full-cascade GB \auc{} of 0.836 versus 0.773 for the local model. A fixed-observation-window or survival analysis would model right-censoring more directly.

\section{Discussion and Threats to Validity}
\label{sec:discussion}

\paragraph{Implications for heterogeneous-data mining}
The pooled co-authorship statistic changes sign when agent composition changes, and omitted context similarly changes the sign of a learned coefficient. Multi-agent operational data should therefore be reported with source-stratified summaries and evaluated under mixture, repository, time, and agent shift. Repository activity must also be computed from PRs preceding the focal record; the paper reports only this historical definition.

\paragraph{Construct validity}
AIDev covers six agents on open-source GitHub from December 2024 through October 2025. The \coauth{} trailer is agent-inserted and nonstandard. More importantly, attribution mode is inferred by matching identity names; unmatched names are called human-attributed even though some agents may use human-looking identities. Missing or asymmetric identities are now marked unresolved, but Codex's 35{,}498-to-10 imbalance across the two analyzed modes still illustrates the risk. We do not claim that this variable measures human effort. Manual validation across agent$\times$mode strata is required before making that interpretation. Task labels are title-derived heuristics, and open PRs are right-censored.

\paragraph{Statistical and causal validity}
The AIPW analysis of the trailer (Section~\ref{sec:robust_assoc}) is observational. Propensity trimming removes many observations, repository-clustered intervals span zero, and unmeasured task difficulty, initial code quality, developer expertise, and governance remain plausible. The observed-mode functional (Section~\ref{sec:ope}) requires adequate overlap and consistent nuisance estimation. Excluding Codex improves overlap but changes the target population, and the bootstrap interval is conditional on fitted nuisance and ranking models. Temporal evaluation retains the context-aware point estimate but does not resolve its small advantage over risk ranking, so it should not be treated as a distinct algorithmic advance.

\paragraph{Decision-time validity}
Only historical repository activity, agent identity, and title-derived task are available when a PR is opened and used for ranking. Attribution mode is constructed from all commits because the snapshot has no commit timestamps. The analysis is therefore retrospective decision support, not a pre-review assignment policy. Consistent with this boundary, the strict creation-time model transfers to unseen repositories (0.710 \auc{}) but degrades under temporal shift (0.641), and temporal evaluation does not distinguish context-aware from risk ranking.

\paragraph{External validity}
The results may not transfer to private repositories, later agent versions, or agents absent from AIDev. LOAO fails for Codex, and Google Jules has only 487 PRs. Future work should validate identities, collect timestamped events, and evaluate a prospective rule under identifiable assignment.

\section{Conclusion}
\label{sec:conclusion}

We have studied 71{,}677 PRs from six AI coding agents as a heterogeneous-data mining problem. Removing Codex changes the pooled co-authorship gap from $-21.8\pp{}$ to $+23.7\pp{}$, while context changes the corresponding logistic-regression coefficient from $-1.09$ to $+1.44$. The source/environment/workflow cascade raises GB \auc{} from 0.774 to 0.832 and retains gains under temporal and unseen-repository evaluation; AIPW intervals show no stable adjusted trailer association.

For the non-Codex population, cross-fitted estimation gives a $+12.3\pp{}$ gain for the 30\% observed-mode functional; the temporal estimate is $+10.3\pp{}$ and indistinguishable from risk ranking ($+0.04\pp{}$). These values standardize name-derived, full-lifecycle metadata states and are not causal human-review effects. Reliable mining of multi-agent operational records therefore requires explicit context and checks for feature timing, mixture shift, and construct validity.

\balance

\end{document}